\documentclass[onecolumn, apj]{emulateapj} 
\usepackage{graphicx}  
\usepackage{dcolumn}   
\usepackage{bm}        
\usepackage{amssymb}   
\usepackage{color}
\usepackage{amsmath}

\def\apj{ApJ}
\def\apjl{ApJL}
\def\apjs{Astrophys. J. Supp. Ser. }

\def\aap{Astron. Astrophys. }

\def\mnras{MNRAS}

\def\icarus{Icarus}

\def\be{\begin{equation}}
\def\ee{\end{equation}}
\def\ba{\begin{eqnarray}}
\def\ea{\end{eqnarray}}
\def\go{\mathrel{\raise.3ex\hbox{$>$}\mkern-14mu
             \lower0.6ex\hbox{$\sim$}}}
\def\lo{\mathrel{\raise.3ex\hbox{$<$}\mkern-14mu
             \lower0.6ex\hbox{$\sim$}}}

\def\tomega{{\tilde{\omega}}}

\begin{document}

\title{Linear Corotation Torques in Non-Barotropic Disks}
\author{David Tsang}\email{dtsang@physics.mcgill.ca} \affiliation{Department of Physics, McGill University, Montreal, QC, Canada}

\date{\today}

\begin{abstract}
I derive a fully analytic expression for the linear corotation torque to first order in eccentricity for planets in non-barotropic protoplanetary disks, taking into account the effect of disk entropy gradients. This torque formula is applicable to both the co-orbital corotation torque and the non co-orbital corotation torques --  for planets in orbits with non-zero eccentricity -- in disks where the thermal diffusivity and viscosity are sufficient to maintain linearity of these interactions. While the co-orbital corotation torque is important for migration of planets in Type I migration, the non co-orbital corotation torque plays an important role in the eccentricity evolution of giant planets that have opened gaps in the disk. The presence of an entropy gradient in the disk can significantly modify the corotation torque in both these cases.  
\end{abstract}
\section{Introduction}

Planets are formed from protoplanetary disks surrounding young stars and are thought to migrate from their birthplaces into their final orbits. Planet-disk interactions tend to dominate the orbital evolution of a planet until the circumstellar disk dissipates after a few Myr. These interactions occur mainly through the exchange of angular momentum and energy around resonances which occur at locations where the natural frequencies of the disk material can be excited by the periodic potential of the planetary perturber \citep{GT79}.

Spiral density waves are launched at Lindblad resonances, located where the radial epicyclic frequency, $\kappa$, matches the frequency of the perturbing potential felt by the disk material. Waves launched from the inner Lindblad resonances tend to drive outward migration of the planet, while waves launched from the outer Lindblad resonances tend to drive inward migration of the planet \citep{GT80}. For small planets within typical protoplanetary disks the outer Lindblad torque dominates over the inner Lindblad torque, and the net differential Lindblad torque drives planet migration inwards, towards the star \citep{Ward86}, though wave reflection may alter the relative strength of the torques near a reflecting edge \citep{Tsang2011}.

For circular orbits,  the corotation resonance --  where the perturbation frequency matches the orbital frequency of the disk material -- occurs at the orbital radius of the planet. For eccentric orbits, non co-orbital corotation resonances can occur both interior and exterior to the planet's semi-major axis. These resonances also exchange energy and angular momentum with the planet. The sign of this exchange depends crucially on the disk parameters at the corotation point. In barotropic disks the sign of the corotation torque depends on the gradient of the vortensity, $\zeta \equiv \omega_z/\Sigma$, where $\omega_z \equiv \hat{z} \cdot ({\bm \nabla} \times {\bm u})$ is the vorticity of the disk, and $\Sigma$ is the disk surface density. This can be understood in a qualitative fashion by considering the effect of the perturbing potential on collisionless particles \citep{GS03}.  

In barotropic disks the vortensity is conserved along streamlines. Material that is slightly outside the corotation point tends to be pushed inwards by the planet's potential, losing angular momentum, while the material slightly inside the corotation tends to be pushed outwards, gaining angular momentum. The corotation torque depends on the relative difference between these two effects, which is determined by preserving the net vortensity at corotation, and thus depends on the value of the vortensity \emph{gradient} at the resonance.  

Many previous simulations of disk torques have utilized a locally isothermal equation of state, and ignore the entropy equation \citep[see e.g.][]{Dunhill2013}. In realistic disks, the barotropic assumption, that pressure depends only on density, does not necessarily hold. In such non-barotropic disks, vortensity is no longer conserved along streamlines, and can be modified by a baroclinic term \citep[see e.g.][]{Lovelace1999}, which arises due to entropy gradients. This baroclinicity also modifies the corotation torque, as it changes the relative difference in the sign of the angular momentum transferred to the disk material in order to preserve the vortensity at the corotation \citep[for a more thorough discussion see Section 2 of][]{Tsang2013b}. 

Previous works on the effect of non-barotropic equations of state on the corotation \citep{BM08, Paardekooper2008, Paardekooper2010, Paardekooper2011} have primarily utilized numerical techniques to study the co-orbital corotation torques. While these works also consider the linear corotation torque, much more emphasis is appropriately placed on the role of non-linear horseshoe torques that dominate the co-orbital region. These non-linear horseshoe torques dominate the migration of small embedded planets, and for typical parameters, the co-orbital corotation torque was found to remain linear only for the order of a libration time \citep{Paardekooper2009, Paardekooper2010} before becoming non-linear, unless diffusivity and viscosity are sufficiently high. Detailed simulations have also been performed \citep[e.g.][]{PaardekooperMellema2008, Kley2008, Bitsch2013a} including these effects for small planets in Type I migration, showing that indeed the effect of entropy gradients on the co-orbital corotation torques can halt or even reverse the migration of embedded planets. 

For planets with small eccentricity, however, the non co-orbital corotation torques are not as likely to become non-linear, as they are located further from planet, and result from potential components that scale linearly with the eccentricity. 

In the barotropic limit, \citet{GT79} calculated the corotation torque for a ``cold disk'', where the disk enthalpy response is negligible compared to the perturbing potential. \citet{Tanaka02} revisited this calculation and numerically calculated the linear enthalpy response of the disk, allowing a semi-analytic calculation of the corotation torque to be provided. \citet{ZL06}, in turn, derived the analytic disk enthalpy response, and were able to provide a fully analytic form for the barotropic corotation torque. 

For non-barotropic disks, \citet{BM08} and \citet{Paardekooper2008} utilized a numerical evaluation of the enthalpy response, similar to \citet{Tanaka02}, to compute a semi-analytic linear corotation torque. Here, we will adopt the approach of \citet{ZL06} and develop a fully analytic solution for the non-barotropic corotation torque. 

In Section 2, I outline the basic equations for a non-barotropic disk. In Section 3, I discuss the evaluation of the torque through the advective angular momentum flux. In Section 4, I analytically compute the enthalpy perturbation at corotation due to a planetary perturber, and in Section 5, use this result to compute a fully analytic expression for the non-barotropic corotation torque. I then discuss the effects of thermal saturation in Section 6 and calculate the necessary thermal diffusivity to maintain linearity. Finally I will summarize the results of this paper in Section 7. 

In a companion paper to this work, \citet{Tsang2013b}, we discuss the effect of non-barotropic corotation torques on the eccentricity evolution 
for giant planets that can clear a gap in the disk. Utilizing results of the calculations in this paper we show that stellar insolation of the gap \citep{Varniere2006, Turner2012, Jang-Condell2012} can result in entropy gradients and eccentricity excitation of giant planets, rather than the damping expected for a barotropic disk. We also suggest that the recently discovered ``Eccentricity Valley'' for low-metallicity exoplanetary systems \citep{Dawson2013} may be a signature of this effect.

\section{Basic Equations}
We begin by limiting ourselves to the examination of two-dimensional perturbations, and proceed with the vertically integrated disk variables
$\Sigma = \int \rho dz$ and $P = \int p dz$, the surface density and vertically integrated pressure respectively. We assume that the unperturbed axisymmetric disk state is given by the general $\Sigma = \Sigma(r)$ and $P = P(r)$ (as opposed to the strictly barotropic case $P = P(\Sigma)$), and the disk velocity profile is (nearly) Keplerian ${\bf u}_o = r\Omega \hat{\phi}$. We take the cylindrical coordinates ($r, \phi, z, t$) to be centered at the central star. 

The continuity and momentum equations are
\ba
\partial_t \Sigma + {\bm u} \cdot \nabla \Sigma + \Sigma \nabla \cdot {\bm u} = 0,\\
\partial_t {\bm u} + (\bm u\cdot \nabla) {\bm u} = - \frac{1}{\Sigma} \nabla P - \nabla \Phi,
\ea
where $\Phi$ is the gravitational potential, including both the star and the planetary perturber. Here we assume that the disk is pressure dominated and will work in the Cowling approximation, assuming the disk self-gravity to be negligible. The star's Newtonian potential is given by $\Phi_s = -(G M_*)/|{\bm r}|$ while the planet at location ${\bm r}_p$ has a potential  given by \citet{GT80}
\be
\Phi_p = -\frac{G M_p}{|{\bm r} - {\bm r}_p|} + \frac{M_p}{M_*} \Omega(r)^2 {\bm r}_p \cdot {\bm r}~.
\ee
where we have defined $M_p$ as the mass of the planet and $M_*$ as the mass of the star. For an eccentric planetary perturber we have the semi-major axis $a$ and eccentricity $e$ such that $\Omega_p \equiv \Omega(a)$, and $e \equiv (r_{\rm max} - r_{\rm min})/2a$. We take the eccentricity of the planet's orbit $e \ll 1$. The radial epicyclic frequency of the planet is given by $\kappa_p \equiv \kappa(a)$ where in general $\kappa(r) \equiv (2\Omega/r) d(r^2 \Omega)/dr$. 

The planet's perturbing potential can be expanded in a Fourier series
\be
\Phi_p = \sum_{l=-\infty}^\infty \sum_{m=0}^\infty \Phi_{l,m} \cos [m\phi + (m\Omega_p + (l-m)\kappa_p)t],
\ee
where to first order in $e$ the non-zero Fourier components are given by \citep{GT80} as
\ba
\Phi_{m,m} &=& -\frac{GM_p}{2a}(2- \delta_{m, 0})[b_{1/2}^m(\beta) - f \beta \delta_{m,1}],\\
\Phi_{m\pm1, m} &=& -\frac{GM_p}{2a}e(2- \delta_{m, 0})\left[\left( \frac{1}{2} \pm \frac{m\Omega_p}{\kappa_p} + \frac{\beta}{2} \frac{d}{d\beta}\right)b^m_{1/2}(\beta) - f \beta \left( \frac{3}{2} - \frac{\kappa_p^2}{2\Omega_p^2} \pm \frac{\Omega_p}{\kappa_p}\right)\delta_{m,1} \right],
\ea
where $f \equiv \Omega_p^2 a^3/(G M_*)$, $\beta \equiv r/a$ is the scaled radius, $\delta_{m,n}$ is the Kronecker delta function and 
\be
b_{1/2}^m(\beta) = \frac{2}{\pi} \int_0^\pi \frac{\cos m\phi d\phi}{(1-2\beta \cos \phi + \beta^2)^{1/2}},
\ee
is the Laplace coefficient. Each of these components has a pattern frequency $\omega_{l, m}  =  m\Omega_{l,m} \equiv m\Omega_p + (l-m)\kappa_p$. 

Assuming the linear perturbations have the form $\delta \propto \exp(im\phi - i \omega_{l, m} t)$, we can write the linear perturbations of the continuity and momentum equations
\ba
-i \tomega \delta \Sigma + \frac{1}{r}\frac{\partial}{\partial r} ( \Sigma r \delta u_r) + \frac{i m }{r}\Sigma \delta u_\phi &=& 0,\label{eq1}\\ 
- i \tomega \delta u_r - 2 \Omega \delta u_\phi &=& - \frac{1}{\Sigma} \frac{\partial}{\partial r} \delta P + \frac{\delta \Sigma}{\Sigma^2} \frac{\partial P}{\partial r} - \frac{\partial \Phi_{l,m}}{\partial r},\label{eq2}\\
-i \tomega \delta u_\phi + \frac{\kappa^2}{2\Omega} \delta u_r &=& -\frac{im}{r}\left( \frac{\delta P}{\Sigma} + \Phi_{l,m} \right)\label{eq3},
\ea
where $\tomega \equiv \omega_{l,m} - m\Omega(r)$ is the perturbation frequency experienced in a frame corotating with the disk and $\delta P$, $\delta \Sigma$, and $\delta {\bm u} \equiv \delta u_r \hat{\bm r} + \delta u_\phi \hat{\bm \phi}$ are the Eulerian perturbations of the pressure, density and velocity respectively.
As in \citet{TL09} we examine adiabatic perturbations of non-barotropic disks, where the Lagrangian pressure and density perturbations are related by
$\Delta \Sigma = \frac{1}{c_s^2}\Delta P$
which relates the Eulerian perturbations as 
\be
\delta \Sigma = \frac{1}{c_s^2}\delta P + \left( \frac{1}{c_s^2} \frac{dP}{dr} - \frac{d\Sigma}{dr}\right)\xi_r = \frac{1}{c_s^2}\delta P - \frac{\Sigma^2 N_r^2}{dP/dr} \frac{i\delta u_r}{\tomega} \label{eq4},
\ee
where $c_s(r) \equiv (\partial P/\partial \Sigma)^{1/2}$ is the adiabatic sound speed and can be a general function of $r$,  $\xi_r = i \delta u_r/\tomega$ is the radial Lagrangian displacement, and $N_r$ is the radial Brunt-V\"ais\"ala frequency defined by
\be
N_r^2 \equiv - \frac{1}{\Sigma^2}\left(\frac{dP}{dr} \right) \left(\frac{1}{c_s^2} \frac{dP}{dr} - \frac{d\Sigma}{dr} \right),
\ee
where we have corrected a typographical sign error in the definition of $N_r^2$ from \citet{TL09}.
Defining the enthalpy perturbation $\delta h \equiv \delta P/\Sigma$ and $D_S \equiv \kappa^2 - \tomega^2 + N_r^2$, and combining \eqref{eq1}-\eqref{eq3} to eliminate $\delta u_\phi$ we obtain
\ba
\frac{\partial}{\partial r}(\delta h + \Phi_{l,m}) &=& \frac{2m\Omega}{\tomega r} (\delta h + \Phi_{l,m}) - \frac{\Sigma N_r^2}{dP/dr} \delta h - \frac{D_S}{\tomega} i \delta u_r,\label{eq5}\\
\frac{\partial}{\partial r}(i \delta u_r) &=& \frac{m^2}{r^2 \tomega^2} (\delta h + \Phi_{l,m}) - \frac{\tomega}{c_s^2}\delta h - \left[-\frac{\Sigma N_r^2}{dP/dr} + \frac{m \kappa^2}{2r \Omega \tomega} + \frac{\partial}{\partial r}(\ln r \Sigma) \right]i \delta u_r.
\ea
Further eliminating $\delta u_r$ we arrive at the second-order inhomogeneous differential equation
\ba
\frac{\partial^2 \delta h}{\partial r^2} &-&\left[\frac{\partial}{\partial r} \ln \left(\frac{D_S}{r\Sigma}\right)\right] \frac{\partial \delta h}{\partial r} - \left[\frac{m^2}{r^2} + \frac{D_S}{c_s^2} + \frac{2m}{r\tomega} \frac{\partial}{\partial r} \ln \left(\frac{\Sigma \Omega}{D_S} \right) \right]\delta h\nonumber \\
&-&\left[ \frac{1}{L_S^2} + \frac{\partial}{\partial r}\left( \frac{1}{L_S}\right) - \frac{1}{L_S} \frac{\partial}{\partial r}\ln \left(\frac{D_S}{r\Sigma} \right)+ \frac{4m\Omega}{r \tomega L_S} - \frac{m^2 N_r^2}{r^2 \tomega^2}\right]\delta h\nonumber\\
&\qquad& =-\frac{\partial^2 \Phi_{l, m} }{\partial r^2}+ \left[\frac{\partial}{\partial r} \ln \left(\frac{D_S}{r\Sigma}\right) - \frac{1}{L_S} \right] \frac{\partial \Phi_{l,m}}{\partial r}\nonumber\\
&\qquad& \qquad+ \left[\frac{m^2}{r^2} + \frac{2m}{r\tomega}\frac{\partial}{\partial r} \ln \left(\frac{\Sigma\Omega}{D_S}\right)  + \frac{2m\Omega}{\tomega r L_S} - \frac{m^2N_r^2}{r^2 \tomega^2}\right] \Phi_{l,m}\,, \label{Mastereqn}
\ea
where 
\be
\frac{1}{L_S} \equiv -\frac{\Sigma N_r^2}{dP/dr},
\ee
is the inverse of the length scale related to the entropy variation in the disk background. 

If the adiabatic index $\gamma = c_s^2 \Sigma/P$ is assumed constant, then we can define the two-dimensional entropy  $S \equiv P/\Sigma^\gamma$, such that
\be
N_r^2 = -\frac{1}{\gamma \Sigma} \frac{dP}{dr}\frac{d \ln S}{dr},  \qquad {\rm and } \qquad  \frac{1}{L_S} = \frac{1}{\gamma} \frac{d\ln S}{dr},
\ee
we recover equation (16) of \citet{BM08}. The left hand side of the equation is the homogeneous equation (8) from \citet{TL09}. In the barotropic disk limit ($N_r^2 \rightarrow 0$, $L_s \rightarrow \infty$) equation (13) from \citet{GT79} is recovered. 
Equation \eqref{Mastereqn} is our master equation describing the vertically integrated perturbations and response of a proto-planetary disk.

\section{Disk Torque and The Advective Angular Momentum Flux}

The torque acting on a disk due to the perturber can be evaluated as the time averaged rate of change of the disk angular momentum (in the vertical direction), while the torque acting on the planet by the disk is equal and opposite to this.
\be
\Gamma_{\rm disk} = -\Gamma_{\rm p \rightarrow d} =- \bigg\langle \frac{d L_{z, {\rm disk}}}{dt} \bigg\rangle,
\ee
where $\langle \ldots \rangle$ denotes the time average over a period, and $L_z$ is the total disk angular momentum in the vertical direction.

The vertical angular momentum areal density $l_z = dL_z/dA$ is given by
\ba
l_z  &=& (\Sigma + \delta \Sigma)({\bm r} \times {\bm u})\nonumber , \\
&=& r^2 \delta \Sigma \Omega + r \Sigma \delta u_\phi + r^2\Sigma \Omega + r\delta \Sigma \delta u_\phi.
\ea
The torque surface density of the planet acting on the disk is then given by the conservation of angular momentum
\ba
\gamma_{p \rightarrow d} &=& \partial_t l_z + {\bm \nabla}\cdot (l_z {\bm u}), \nonumber \\
&=& \frac{\partial l_z}{\partial t} + \frac{1}{r}\frac{\partial}{\partial r} ( r l_z \delta u_r) + \frac{1}{r}\frac{\partial}{\partial \phi} [l_z (r\Omega + \delta u_\phi)].
\ea
We then have the total torque on the disk given by
\be
\Gamma_{p\rightarrow d} = \bigg\langle \iint\limits_{\rm disk} dr d\phi \left(\frac{\partial}{\partial t}(r l_z) + \frac{\partial}{\partial r} ( r l_z \delta u_r) + \frac{\partial}{\partial \phi} [l_z (r\Omega + \delta u_\phi)]\right) \bigg\rangle.
\ee
Let us assume that the disk is in a steady state, relative to the orbital timescale, such that all transient perturbations have died away, and we can therefore take $\omega$ to be real.  We see that the first term in the integrand above has zero net contribution when integrated over a period for the time average. Similarly the last term also has no net contribution, after integration over the azimuthal angle $\phi$ from $0$ to $2\pi$. We thus only have a contribution from the second term, which corresponds to the angular momentum flux, 
\be
\Delta F_{\rm L} \equiv \bigg\langle \int_0^{2\pi} d\phi \left( r^2\Sigma\delta u_\phi \delta u_r + r^3\Omega \delta\Sigma \delta u_r + r^3 \Sigma \Omega \delta u_r \right) \bigg\rangle\bigg|_{r_-}^{r_+} ,
\ee
where we are taking the real part of all perturbations, $r_+$ and $r_-$ are the cylindrical boundaries of the part of the disk in which we are interested, and we have kept only up to quadratic order in the perturbation. The last term above is linear in the perturbation and thus yields no contribution when integrated over $\phi$ and averaged over period. 

Expressing the flux in terms of complex perturbations we then have
\ba
F_{\rm L} &=& \bigg\langle \int_0^{2\pi} d\phi \left( r^2 \Sigma \,{\rm Re}[\delta u_r]\, {\rm Re}[\delta u_\phi] + r^3 \Omega\, {\rm Re}[\delta \Sigma]\, {\rm Re}[\delta u_r]\right) \bigg\rangle\\
&=& \pi r^2\Sigma\, {\rm Re}[\delta u_r\, \delta u_\phi^*] + \pi r^3 \Omega\, {\rm Re}[\delta \Sigma\, \delta u_r^*], \label{fluxeqn}
\ea
where ${\rm Re}[z]$ is the real part of $z$, and  $z^*$ denotes the complex conjugate of $z$.

The second term above $F_{\rm mf}(r) \equiv \pi r^3 \Omega\, {\rm Re}[\delta \Sigma\, \delta u_r^*]$ corresponds to the angular momentum transport due to total mass flux through a radius $r$. However, the angular momentum content of this mass is not changing, merely being transported outwards. Therefore, when evaluated over the entire disk $\Delta F_{\rm mf}$ does not provide a contribution to the total torque on the planet. 
Thus the only term that contributes to the torque on the planet is $F_{\rm adv} \equiv \pi r^2\Sigma\, {\rm Re}[\delta u_r\, \delta u_\phi^*]$, the advective angular momentum flux \citep{LK72},

\be
\Gamma_{\rm disk} = -\Gamma_{p \rightarrow d} = -\Delta F_{\rm adv} = F_{\rm adv}(r_-) - F_{\rm adv}(r_+).
\ee

Utilizing \eqref{eq2}-\eqref{eq4}, $F_{\rm adv}$ above can be evaluated for a particular $l, m$ component, in terms of $\delta h$ and $\Phi_{l,m}$
\ba
F_{\rm adv} &=& \frac{m \pi \Sigma r}{D_S}{\rm Im}[(\delta h + \Phi_{l,m})(\delta h' + \Phi'_{l,m})^*] + \frac{1}{L_S} \frac{m\pi \Sigma r}{D_S} {\rm Im}[\delta h\, \Phi_{l,m}^*], \label{flux1}
\ea
\citep{BM08} where ${\rm Im}[z]$ is the imaginary part of $z$ and $f'$ denotes $\partial f/\partial r$. In the barotropic limit $L_S \rightarrow \infty$, and this reduces to the flux from \citet{GT79}.



The torques above can be evaluated through the advective fluxes, either outside the Lindblad resonances, or on either side of the corotation resonance. Note that we do not find a torque contribution from the singularity due to entropy advection in Equation \eqref{eq4}, contrary to \citet{BM08}, who added an extraneous contribution due to the singular entropy perturbation. This was later shown to be due to non-linear effects (thermal saturation) on the corotation torque \citep{Paardekooper2008}, which we will discuss in \S \ref{nonlinear}.

The torque expression above requires the enthalpy perturbation as a function of the forcing potential, particularly near the corotation resonance. \citet{BM08} and \citet{Tanaka02} approach this semi-analytically, by leaving this enthalpy perturbation (and its derivative) in the expression and computing it numerically. Here we will adopt the approach of \citet{ZL06}, where we will explicitly solve the disk enthalpy response to the forcing potential analytically near the resonances to find the torque. 

\section{Corotation Resonance}

Expanding equation \eqref{Mastereqn} near the corotation resonance, where $\tomega(r_c) = 0$, and keeping only singular terms and those of order $\sim (H/r)^{-2} \equiv r^2\Omega^2/c_s^2$, where $H$ is the disk scale height, we find
\be
\frac{d^2 w}{dr^2} -\left[\frac{D_S}{c_s^2} + \frac{2}{q} \left(\frac{d}{dr} \ln \left(\frac{\Sigma \Omega}{D_S} \right) + \frac{2}{L_S}\right)\frac{1}{r-r_c}- \frac{N_r^2}{q^2 \Omega^2} \frac{1}{(r-r_c)^2}\right]w  =-\left[\frac{D_S}{c_s^2} + \frac{2}{qL_S}\frac{1}{r-r_c}\right]\Phi_{l,m}, \label{singequation}
\ee
where $w \equiv \delta h + \Phi_{l,m}$, and $q \equiv -( d\ln \Omega/d\ln r)_{r_c}$ and we've assumed that $m^2/r^2 \ll 1/H^2$. 

Utilizing the Landau prescription to avoid the singularities by taking $z \equiv x + i\epsilon$, for some small $\epsilon > 0$ (such that $-\pi \leq {\rm Arg}(z) \leq 0$) where  $x \equiv \int_{r_c}^r 2 k dr$, and defining $k^2 \equiv D_S/c_s^2$, and $\psi \equiv k^{1/2}w$ we can further simplify 
\be
\frac{d^2}{dz} \psi + \left[-\frac{1}{4} + \frac{\nu}{z} + \frac{1/4 - \mu^2}{z^2} \right]\psi = -\left[\frac{1}{4} + \frac{c_s}{q \kappa L_S}\frac{1}{z}\right]\frac{1}{\sqrt{k}}\Phi_{l,m} \label{whittaker},
\ee
where we have assumed that $k$ does not change quickly near the corotation such that $D_S/c_s^2 \gg k^{1/2}\, \partial_r^2 (k^{-1/2})$, and defined the important parameters 
\be
\nu \equiv \frac{c_s}{q\sqrt{D_S}} \left(\frac{d}{dr}\ln\zeta - \frac{2}{L_S} \right) \bigg|_{r_c},\qquad  {\rm and}  \qquad \mu \equiv \frac{1}{2}\left(1 - \frac{4 N_r^2}{q^2 \Omega^2} \right)^{1/2}\bigg|_{r_c},
\ee
where $\zeta \equiv D_S/(2\Sigma\Omega)$ evaluated at the corotation is the vortensity for the barotropic case.
We recognize the homogeneous version (when $\Phi_{l,m} = 0$) of equation \eqref{whittaker} as the Whittaker differential equation \citep{AS65, BatemanProject, Olver:2010:NHMF}, which is solved, unsurprisingly,  by the Whittaker function $W_{\nu, \mu}(z)$. We can choose the two linearly independent homogenous solutions to be 
\be
k^{1/2} w_1 = \psi_1 \equiv {\rm W}_{\nu,\mu}(z) \qquad {\rm and} \qquad k^{1/2} w_2 = \psi_2 \equiv {\rm W}_{-\nu, \mu} (z\,e^{i\pi})
\ee
such that the asymptotic forms are convenient for determining the boundary conditions\footnote{For a more thorough discussion of asymptotic expansion and connection formulae for Whittaker functions involving Stokes phenomenon see Appendix A in \citet{Tsang08}.}. These are given as 
\ba
w_1 &\sim&
\Biggl\{\begin{array}{ll}
\tfrac{1}{\sqrt{k_{\rm eff}}}\, \exp\left(-\int_{r_c}^{r}\! k_{\rm eff} \,dr \right) 
& \qquad \qquad  ~~\textrm{for } r \gg r_{c},\\
\tfrac{1}{\sqrt{k_{\rm eff}}}\, e^{i\pi\nu}\exp\left(+\int_r^{r_{c}}\! k_{\rm eff} \,dr \right) 
+ \tfrac{1}{\sqrt{k_{\rm eff}}}\, \frac{T_1}{2} e^{-i\pi\nu} \exp\left(-\int_r^{r_{c}}\! k_{\rm eff} \,dr \right) 
&\qquad \qquad~~\textrm{for } r\ll r_{c},
\end{array}\label{whitaker1}\\
w_2 &\sim&
\Biggl\{\begin{array}{ll}
\tfrac{1}{\sqrt{k_{\rm eff}}}\, e^{-i\pi\nu} \exp\left(+\int_{r_c}^{r}\! k_{\rm eff} \,dr \right) + \tfrac{1}{\sqrt{k_{\rm eff}}}\tfrac{T_1}{2}e^{i\pi\nu} \exp\left(-\int_{r_c}^{r}\! k_{\rm eff} \,dr \right)
&\qquad \qquad ~~~~ \textrm{for } r \gg r_{c},\\
\tfrac{1}{\sqrt{k_{\rm eff}}}\, e^{-i2\pi\nu} \exp\left(-\int_r^{r_{c}}\! k_{\rm eff} \,dr \right) &\qquad \qquad ~~~~ \textrm{for } r\ll r_{c},
\end{array}\label{whitaker2}
\ea
where the Stokes multipliers (Heading 1962) are given by the $\Gamma$ function, 
\be 
T_0={2\pi i\over\Gamma(\tfrac{1}{2} - \mu + \nu)\Gamma(\tfrac{1}{2} + \mu+\nu)},\qquad
T_1={2\pi i\,e^{i2\pi\nu}\over\Gamma(\tfrac{1}{2} - \mu -\nu)\Gamma(\tfrac{1}{2} + \mu -\nu)},\label{stokes}.
\ee
and $\sim$ denotes an asymptotic expansion and $\gg$ and $\ll$ are here taken to mean the range of validity for an asymptotic expansion of a local solution.

We see above that the solutions $w_1$ and $w_2$ decay exponentially away from the corotation for $r \gg r_c$ and $r \ll r_c$ respectively, while both are unbounded on the opposing sides of the corotation. 

The solution to the inhomogeneous Equation \eqref{whittaker} can then be given by the method of variation of parameters to be
\ba
w(x) &=& -{\rm W}_{\nu,\mu}(z) \int_{-\infty}^{x} \frac{{\rm W}_{-\nu, \mu}(-z)}{\cal W} \left(-\frac{1}{4} -\frac{c_s}{q\kappa L_S}\frac{1}{z} \right)\Phi_{l,m} dx \nonumber \\ 
&\qquad&~~ - {\rm W}_{-\nu,\mu}(-z) \int_{x}^{\infty} \frac{{\rm W}_{\nu,\mu}(z)}{\cal W}\left(-\frac{1}{4} -\frac{c_s}{q\kappa L_S}\frac{1}{z} \right)\Phi_{l,m} dx, \label{varofparam}
\ea
where ${\cal W} \equiv \psi_1 \psi_2' - \psi_1' \psi_2 = e^{-i \pi \nu}$ is the Wronskian \citep{Olver:2010:NHMF}, and we have taken the limits of integration such that the solution is bounded on either side of the corotation.

For the calculation of the torque we will need to evaluate Equation \eqref{varofparam} at $x=0$. To do this we can utilize the Laplace transform identity for Whittaker functions \citep{BatemanProject, Olver:2010:NHMF},
\be
\int_0^\infty e^{-st} t^{b - 1} {\rm W}_{\nu, \mu}(t) dt = \frac{\Gamma(\tfrac{1}{2} + \mu + b)\, \Gamma(\tfrac{1}{2} - \mu + b)}{\Gamma(1 - \nu + b)}\, _2{\rm F}_1(\tfrac{1}{2} -\mu + b, \tfrac{1}{2} + \mu + b; 1-\nu + b; \tfrac{1}{2} - s), \label{Laplace}
\ee
which is valid for ${\rm Re}[\mu] < {\rm Re}[b + \tfrac{1}{2}]$ and ${\rm Re}[s] > -\tfrac{1}{2}$, and where $_2{\rm F}_1(a, b;c;z)$ is the Gaussian (ordinary) hypergeometric function. This gives us
\be
\int_0^\infty  {\rm W}_{\nu,\mu}(x)dx =  \frac{\Gamma(\tfrac{3}{2} + \mu)\,\Gamma(\tfrac{3}{2}-\mu)}{\Gamma(2 - \nu)}\,_2{\rm F}_1(\tfrac{3}{2} - \mu, \tfrac{3}{2} + \mu; 2-\nu; \tfrac{1}{2}),
\ee
for $b = 1$ and $s = 0$, and
\be
\int_0^\infty \frac{1}{x} {\rm W}_{\nu,\mu}(x)dx =  \frac{\Gamma(\tfrac{1}{2} + \mu)\,\Gamma(\tfrac{1}{2}-\mu)}{\Gamma(1 - \nu)}\,_2{\rm F}_1(\tfrac{1}{2} - \mu, \tfrac{1}{2} + \mu; 1-\nu; \tfrac{1}{2}),
\ee
for $b=0$ and $s =0$, where we note that the Laplace transform \eqref{Laplace} is valid to use since $1/2 - \mu > 0$ for radially stable stratified disks\footnote{Note that it is possible for some disks to have regions where $N_r^2 < 0$, for instance, those where the nonlinear baroclinic instability has only a slow growth rate \citep{Lesur2010}, or where magnetic field has stabilized the disk \citep{Lyra2011}. In these locations a numerical calculation of the disk response should be used, though it may be possible to extend the analytic analysis above.} with $N_r^2 > 0$. 

This allows us to evaluate Equation \eqref{varofparam} at $r = r_c$, ($x = 0$), assuming all disk properties and the forcing potential to be roughly constant in the vicinity of the corotation,  
\ba
w(r_c) &=& \frac{\Phi_{l,m}}{4}e^{i\pi \nu} \bigg[{\rm W}_{\nu,\mu}(i\epsilon)\,\frac{\Gamma(\tfrac{3}{2} + \mu)\,\Gamma(\tfrac{3}{2}-\mu)}{\Gamma(2 + \nu)} \,_2{\rm F}_1(\tfrac{3}{2} - \mu, \tfrac{3}{2} + \mu; 2+\nu; \tfrac{1}{2})\nonumber\\ 
&\qquad& \qquad \qquad ~~~~ + {\rm W}_{-\nu,\mu}(-i\epsilon)\,\frac{\Gamma(\tfrac{3}{2} + \mu)\,\Gamma(\tfrac{3}{2}-\mu)}{\Gamma(2 - \nu)} \,_2{\rm F}_1(\tfrac{3}{2} - \mu, \tfrac{3}{2} + \mu; 2-\nu; \tfrac{1}{2}) \bigg]\nonumber\\
&\qquad&   - \frac{c_s \Phi_{l,m}}{q\kappa L_S} e^{i\pi\nu} \bigg[{\rm W}_{\nu,\mu}(i\epsilon)\,\frac{\Gamma(\tfrac{1}{2} + \mu)\,\Gamma(\tfrac{1}{2}-\mu)}{\Gamma(1 + \nu)}\,_2{\rm F}_1(\tfrac{1}{2} - \mu, \tfrac{1}{2} + \mu; 1+\nu; \tfrac{1}{2})\nonumber \\ 
&\qquad& \qquad \qquad \qquad~~~  - {\rm W}_{-\nu,\mu}(-i\epsilon)\,\frac{\Gamma(\tfrac{1}{2} + \mu)\,\Gamma(\tfrac{1}{2}-\mu)}{\Gamma(1 -\nu)}\, _2{\rm F}_1(\tfrac{1}{2} - \mu, \tfrac{1}{2} + \mu; 1-\nu; \tfrac{1}{2}) \bigg]. \label{w_at_rc}
\ea

For general small imaginary component $\epsilon > 0$ from the Landau prescription above, we can evaluate ${\rm W}_{\pm\nu,\mu}(\pm i\epsilon)$ using Equation \eqref{Wnearzero}. We note that formally for $\mu \neq 1/2$ as $|z|\rightarrow 0$ then  ${\rm W}_{\nu, \mu}(z) \rightarrow 0$ , however, it does so only logarithmically slowly.  Thus as we show in Appendix \ref{WhittZero}, for sufficient large (but still quite small) $\epsilon$, we can approximate 
\be
{\rm W}_{\nu,\mu}(i\epsilon) \simeq {\rm W}_{\nu, 1/2}(0)  = \frac{1}{\Gamma(1-\nu)},
\ee
which gives the value for $w = \delta h + \Phi_{l,m}$ evaluated at the corotation, 
\ba
w(r_c) &\simeq& \frac{\Phi_{l,m}(r_c)}{4}e^{i\pi \nu} \frac{\tfrac{1}{4} - \mu^2}{\sin \pi(\tfrac{1}{2} - \mu)} \frac{\sin \pi \nu}{\nu}  \bigg[\frac{{\cal F}(\mu, \nu)}{1 + \nu}
+  \frac{{\cal F}(\mu, -\nu)}{1-\nu} \bigg]\nonumber\\
&\qquad& - \frac{c_s \Phi_{l,m}(r_c)}{q\kappa L_S} e^{i\pi\nu} \frac{1}{\sin \pi(\tfrac{1}{2} - \mu)}\frac{\sin \pi \nu}{\nu} \bigg[{\cal G}(\mu, +\nu)
-{\cal G}(\mu, -\nu) \bigg]\label{approxw},
\ea
where we have utilized the well known recurrence relation, $z\Gamma(z) = \Gamma(1 + z)$, and reflection formula, $\Gamma(1-z)\Gamma(z) = \pi/\sin(\pi z)$ for the gamma function and we have defined
\be
{\cal F}(\mu,\nu) \equiv \,_2{\rm F}_1(\tfrac{3}{2} - \mu, \tfrac{3}{2} + \mu; 2+\nu; \tfrac{1}{2}),\qquad {\rm and} \qquad  
{\cal G}(\mu,\nu) \equiv \,_2{\rm F}_1(\tfrac{1}{2} - \mu, \tfrac{1}{2} + \mu; 1+\nu; \tfrac{1}{2}).
\ee
In the barotropic limit we have $(\tfrac{1}{2} - \mu) \sim N_r^2 \rightarrow 0$ and $L_S^{-1} \sim N_r^{2} \rightarrow 0$, and noting that ${\cal G}(1/2, \pm\nu)  = 1$ we can recover Equation (107) from \citet{ZL06} (where their $p/2q = \nu$).

\section{The Non-Barotropic Corotation Torque}
We can arrive at an expression for the linear non-barotropic corotation torque by dividing Equation \eqref{singequation} by $w$ (which we have shown in Appendix \ref{WhittZero} to be non-zero at corotation for small, but finite imaginary component $z = i\epsilon$) and then integrating over the singularity at corotation,
\be
\frac{1}{w(r_c)} \frac{dw}{dr}\bigg|_{r_c^-}^{r_c^+} = + i \pi \frac{2}{q}\left( \frac{d}{dr}\ln \zeta - \frac{2}{L_S}\right)  + i\pi \frac{2}{qL_S}\frac{\Phi_{l,m}(r_c)}{w(r_c)},
\ee
where we have evaluated across the residues assuming $-\pi \leq {\rm Arg}(r-r_c) \leq 0$ consistent with the Landau prescription above. This gives the discontinuity at the corotation for the first derivative of $w$,
\be
\Delta w'(r_c) = w'(r_c^+) - w'(r_c^-) = i\pi \frac{2}{q}\left( \frac{d}{dr} \ln \zeta -\frac{2}{L_S}\right) \delta h(r_c) + i\pi \frac{2}{q}\left( \frac{d}{dr} \ln \zeta -\frac{1}{L_S}\right) \Phi_{l,m}(r_c).
\ee
Equation \eqref{flux1} can then be evaluated on either side of the corotation as
\be
\Delta F_{\rm adv} = \frac{m\pi \Sigma r}{D_S} {\rm Im}\left[w(r_c) \Delta w'(r_c) \right] + \frac{1}{L_S} \frac{m\pi\Sigma r}{D_S}{\rm Im}[\delta h \Phi_{l,m}^*]\bigg|_{r_c^-}^{r_c^+}.
\ee
The second term above is clearly zero across corotation for the non co-orbital corotation resonances $l = m\pm 1$, as at these distinct locations the potential components $\Phi_{m\pm 1, m}$ are continuous. For the co-orbital resonances ($l = m$), the potential components, $\Phi_{m,m}$, diverge logarithmically at corotation,  however the inclusion of an arbitrarily small softening length will regularize the potential at corotation \citep{BM08}, and we may take $\Phi_{m,m}$ to be continuous, and thus the second term above to again be zero.

Combining the above equations we have 
\be
\Gamma_{\rm disk} = -\Delta F_{\rm adv} = -\frac{m\pi^2 \Sigma r}{D_S}\frac{2}{q} \left[\left(\frac{d}{dr} \zeta - \frac{2}{L_S} \right) |w(r_c)|^2  + \frac{1}{L_S} \Phi_{l,m} {\rm Re}[w(r_c)]\right],
\ee
where we have assumed a phase such that $\Phi_{l,m}$ is purely real.
Utilizing Equation \eqref{approxw} we then find an expression for the linear corotation torque on the planet due to the disk,

\be
\begin{split}
\Gamma^{\rm (CR)}_{l,m} &=  -\left[ \frac{2m \pi^2 \Sigma \Phi_{l,m}^2}{(d\ln\Omega/dr)D_S} \left(\frac{d}{dr} \ln \zeta - \frac{2}{L_S}\right)\right]_{r_c} 
\times \Bigg\{  \frac{\pi(\tfrac{1}{4} - \mu^2)}{\sin \pi(\tfrac{1}{2}-\mu)}\frac{\sin\pi\nu}{\pi\nu}  
\frac{1}{4}\left(\frac{{\cal F}(\mu, \nu)}{1 + \nu} +  \frac{{\cal F}(\mu, -\nu)}{1-\nu} \right)\\
&\qquad \qquad\qquad\qquad\qquad \qquad \qquad \qquad \quad \qquad  - \left(\frac{c_s}{q\sqrt{D_S}}\frac{1}{L_S}\right)_{r_c} \frac{\pi}{\sin \pi(\tfrac{1}{2} - \mu)}\frac{\sin \pi \nu}{\pi \nu}
 \left[{\cal G}(\mu, \nu) - {\cal G}(\mu,-\nu)\right]
 \Bigg\}^2\\
&\qquad - \left[\frac{2m \pi^2 \Sigma \Phi_{l,m}^2}{(d\ln\Omega/dr)D_S} \frac{1}{ L_S} \right]_{r_c} \cos(\pi \nu)
\times \Bigg\{  \frac{\pi(\tfrac{1}{4} - \mu^2)}{\sin \pi(\tfrac{1}{2}-\mu)}\frac{\sin\pi\nu}{\pi\nu}  
\frac{1}{4}\left(\frac{{\cal F}(\mu, \nu)}{1 + \nu} +  \frac{{\cal F}(\mu, -\nu)}{1-\nu} \right)\\
&\qquad \qquad\qquad\qquad\qquad \qquad \qquad \qquad \quad \qquad  - \left(\frac{c_s}{q\sqrt{D_S}}\frac{1}{L_S}\right)_{r_c} \frac{\pi}{\sin \pi(\tfrac{1}{2} - \mu)}\frac{\sin \pi \nu}{\pi\nu}
 \left[{\cal G}(\mu, \nu) - {\cal G}(\mu,-\nu)\right]
 \Bigg\}. \label{torqueeqn}
\end{split}
\ee
 Taking the isentropic limit $N_r^2 \rightarrow 0$, such that $1/L_S \rightarrow 0$, $\mu \rightarrow 1/2$, and $\pi(\tfrac{1}{4} - \mu^2)/(\sin\pi[\tfrac{1}{2} - \mu]) \rightarrow 1$, recovers Equation (109) from \citet{ZL06}, and further taking the `cold disk' limit of $\nu \sim c_s/(r\Omega) \rightarrow 0$ (and noting that ${\cal F}[1/2, 0] = 2$) we obtain the classical barotropic cold-disk corotation torque from \citet{GT79}. 

Taking the cold disk limit, $c_s/(r\Omega) \rightarrow 0$, such that $\sin \pi \nu/(\pi \nu) \rightarrow 1$ and ${\cal G}(\mu, \nu) - {\cal G}(\mu, -\nu)\rightarrow 0$,  but not the isentropic limit such that $N_r^2 > 0$, we obtain
\ba
\Gamma^{\rm (CR, cold)}_{l,m} &=& -\left[ \frac{2m \pi^2 \Sigma \Phi_{l,m}^2}{(d\ln\Omega/dr)D_S}\right]_{r_c} \left\{
\left(\frac{d}{dr} \ln \zeta - \frac{2}{L_S}\right) \times \left[  \frac{\pi(\tfrac{1}{4} - \mu^2)}{\sin \pi(\tfrac{1}{2}-\mu)}
\frac{{\cal F}(\mu, 0)}{2} \right]^2 +  \frac{1}{ L_S} 
\times \left[ \frac{\pi(\tfrac{1}{4} - \mu^2)}{\sin \pi(\tfrac{1}{2}-\mu)}
\frac{{\cal F}(\mu, 0)}{2} \right]\right\},\nonumber \\
&=& -\left[\frac{2 m \pi^2 \Sigma \Phi_{l,m}^2}{(d\ln\Omega/dr) \kappa^2} \right]_{r_c}\left\{\left(\frac{d}{dr} \ln \zeta - \frac{1}{L_S}\right)+ {\cal O}(N_r^4/\Omega^4) \right\}
\ea
where the second equality is true in the limit where $|r/L_S| \sim |d\ln\Sigma/d\ln r| < (H/r)^{-1}$. The error in using this cold disk limit in the above equation for small $N_r^2 > 0$ scales with ${\cal O}(N_r^4/\Omega^4)$. This simplified cold-disk limit for the non-barotropic torque is used in a companion paper to this work, \citep{Tsang2013b}, which discusses the impact of the non-barotropic torque on eccentricity evolution of giant planets.  When $L_S \sim 1/(d\ln\Sigma/dr) \sim H$ then $N_r^2 \sim \Omega^2$ and the full form of \eqref{torqueeqn} should be used. 

\section{Non-Linear Saturation and Thermal Diffusivity}\label{nonlinear}

Three separate non-linear effects can play a role in the evolution of the corotation torque in non-barotropic disks. The first is due to the the singular density and entropy perturbation at corotation \citep{Paardekooper2008}, which can cause the thermal effect to become saturated, reducing the torque to that of the barotropic case. The second is the onset of the non-linear horseshoe torque in the co-orbital region \citep{Ward1991}, which dominates over the linear corotation torque after a libration timescale \citep{Paardekooper2009, Paardekooper2010}, though this does not occur for the non co-orbital resonances. The third is the saturation of the corotation region, where the background quantities have been sufficiently modified by the corotation interaction to reduce or halt the corotation torque entirely \citep{Paardekooper2009, Paardekooper2010}. In all cases sufficient viscosity and/or thermal diffusivity can prevent the non-linear effects from arising, and the linear corotation torque we have derived will remain valid \citep{Paardekooper2011}. 

Here we will consider in detail the effect of the singular entropy perturbation, as this can quickly saturate the effects of the entropy gradient, returning the corotation torque to its barotropic value. This occurs at both the co-orbital ($l=m$) and non co-orbital ($l = m \pm 1$) corotation resonances. However, following the discussion of \citet{Paardekooper2008} we will show that sufficient thermal diffusivity restores the linear non-barotropic torque. 

The linear entropy perturbation can be defined as
\be
\frac{\delta S}{S} \equiv \frac{\delta P}{P} - \gamma \frac{\delta \Sigma}{\Sigma},
\ee
which, when combined with equation Equation \eqref{eq4} yields
\be
\delta S = - \frac{\partial S}{\partial r} \frac{i \delta u_r}{\tomega}.
\ee
which corresponds to the singular density perturbation at the same location. In the non-dissipative non-diffusive limit in which we have performed our linear calculations these singular density and entropy perturbations result in non-linear effects that arise as the perturbation amplitude grows large \citep{Paardekooper2008}. When $\partial_r \delta S \sim dS/dr$ the gradient of the entropy due to the perturbation is comparable to the background entropy gradient, and the corotation can thermally saturate, reducing the torque to the barotropic value. This effect, seen in simulations, was misattributed to a component of the linear torque due to the singular density perturbation by \citet{BM08}, however, this singular component  does not contribute directly to the linear corotation torque. 

Including the effect of thermal diffusion the above equation can be rewritten
\be
-i\tomega \delta S + \delta u_r \frac{dS}{dr} = \frac{K}{\rho C_p} \nabla^2 \delta S,
\ee
where $\rho$ is the volume density, $K$ is the thermal conductivity and $C_p$ is the specific heat at constant pressure, and where we have ignored the effect of the pressure perturbation which is well behaved at the corotation.

Expanding around the corotation \citet{Paardekooper2008} showed that the equation above can be rewritten in terms of the inhomogeneous Airy differential equation, which can be solved (assuming bounded behavior away from the corotation) by the inhomogeneous Airy function  ${\rm Hi}(z)$ \citep{AS65, Olver:2010:NHMF}, such that
\be
\delta S = - \left( \frac{3^{1/3} \pi F}{\lambda^{2/3}}\right) {\rm Hi}(\xi),
\ee
where $\lambda \equiv 3 m /(2r_c^3 D_e)$, $D_e\equiv (2 H K)/(\Sigma C_p r_c^2 \Omega_c) = (2 c_s K)/(\Sigma C_p r_c^2 \Omega^2)$ is the dimensionless diffusivity, $F \equiv \delta u_r (dS/dr)/(D_e r_c^2 \Omega_c)$, $\Omega_c \equiv \Omega(r_c)$ and
\be
\xi = \frac{2(i\tomega - m^2 D_e \Omega_c)(9m/2D_e)^{1/3}}{3m\Omega_c}.
\ee
This manifests as an entropy peak located at the corotation, with the perturbation amplitude given by $|\delta S| \sim  3^{1/3} \pi F/ \lambda^{2/3}$, and the length scale of entropy perturbation at corotation is then given by $\Delta r_S \sim \lambda^{-1/3}$, which implies that the corotation becomes thermally saturated when the background entropy gradient is comparable to that of the perturbation, 
\be
\frac{dS}{dr} \sim \frac{|\delta S(r_c)|}{\Delta r_S} \sim  \frac{dS}{dr} \frac{1}{D_e r_c^2 \Omega_c} \left(\frac{r_c^3 D_e}{m}\right)^{1/3} |\delta u_r|,
\ee
thus non-barotropic linearity is preserved when the dimensionless diffusion constant is greater than
\be
D_e \gg \frac{1}{m^{1/2}} \left(\frac{|\delta u_r|}{r \Omega} \right)_{r_c}^{3/2} =  m\left(\frac{2 |w(r)|}{r^2\kappa^2}\right)_{r_c}^{3/2}.
\ee
where we have utilized Equation \eqref{eq5} evaluated at the corotation. From Equation \eqref{approxw}, we see we can estimate $|w(r_c)| \sim \Phi_{l,m}$ for small $\nu$ and $1/2 - \mu$, which allows us to estimate the thermal diffusivity necessary to maintain linearity for the non co-orbital corotation resonance,
\be
D_e \gg m \left(\frac{|\Phi_{m\pm1,m}|}{r^2 \kappa^2}\right)_{r_c}^{3/2} \simeq m  \left|\left(e\frac{M_p}{M_*}\beta\right)\times \left(1 \pm \frac{2m\Omega_p}{\kappa_p} + \beta \frac{d}{d\beta}\right) b_{1/2}^m(\beta) \right|_{r_c}^{3/2} \sim e^{3/2} \left( \frac{M_p}{M_*}\right)^{3/2},
\ee
while for the co-orbital corotation the diffusivity required depends on the amount of softening used for the potential, as well as the perturbations caused by other azimuthal components of the potential. \citet{Paardekooper2008} numerically found that $D_e > 10^{-6} - 10^{-5}$ was sufficient to prevent thermal saturation for the co-orbital torques in their simulations of smaller earth-mass planets. 
Explicitly evaluating the above expression for a planet located at 5 AU \citep[the example discussed in][]{Tsang2013b,Turner2012},  in a Keplerian disk we see that the $m=3$ non co-orbital corotation resonance located at $\simeq 6.06 $ AU remains thermally unsaturated if  $D_e \gg 6.3 \times 10^{-8} (e/10^{-2})^{3/2} [M_p/(10^{-3} M_*)]^{3/2}$. Including thermal diffusion with $D_e$ obeying these conditions is equivalent to adding a small positive $\epsilon$ in the Landau prescription.

\section{Discussion and Conclusion}
We have examined the linear torque due to planet-disk interaction at the corotation resonance in non-barotropic disks. 
While other works have previously provided semi-analytic expressions that required numerical evaluation and numerical fitting formula \citep{BM08, Paardekooper2008, Paardekooper2010} to the co-orbital corotation torque, we have developed a fully analytic expression for the linear corotation torque in a non-barotropic disk, for both the co-orbital and non co-orbital corotation resonances. The main result of this work is Equation \eqref{torqueeqn}, which generalizes the corotation torque expressions of \citet{ZL06} and \citet{GT79}.

For small planets, which have not cleared a gap in the disk, co-orbital corotation torque is likely to be linear only for roughly a libration time \citep{Paardekooper2009, Paardekooper2010}, and the non-linear horseshoe torque will likely develop. However, the linearity of the system is more easily preserved for the non co-orbital corotation resonance, as the potential varies much more smoothly far from the location of the planet, and the perturbing potential is additionally reduced by a factor of the eccentricity $e$. For higher $m$, the non co-orbital corotation resonances are more closely bunched, and interactions between resonances can become important depending on the disk properties. For larger planets that have sufficiently clean gaps, the higher-$m$ resonances located within the gap have their torques suppressed due to the low surface density. We also note that the torque formula we have derived is not valid for $m \go r/H$, above which the torque is cutoff due to the high resonance order and finite thickness effects \citep{Ward1989}.

We have shown that entropy gradients in the disk, whether they arise from disk heating, opacity changes, or stellar illumination, can significantly modify the corotation torque acting on planets. Additionally, we have also calculated the minimum thermal diffusivity required to prevent thermal saturation, below which the torque returns to its barotropic value. 

We have used a general prescription for adiabatic perturbations and have not assumed a particular form for the equation of state or energy equation, expressing both the radial  Brunt-V\"ais\"ala frequency, $N_r$, and the characteristic entropy length scale $L_S$ in terms of density and pressure gradients. For simplified models these key quantities can also be expressed in terms of the entropy $S$ and the adiabatic index $\gamma$ for particular equations of state. In taking the perturbations to be adiabatic we have implicitly assumed that the characteristic timescale for energy transport in the disk is slower than the perturbation (orbital) timescale, as is the case for radiatively inefficient disks.

The expression for the corotation torque in Equation \eqref{torqueeqn} above should be applicable to both Keplerian disks, where the pressure gradients are negligible, and to disks that are significantly non-Keplerian, due to either the effects of general relativity, or strong pressure gradients \citep[see e.g.][]{GS03}.  We note that for truly sharp density or pressure transitions, such as gaps that have edge transitions of order less than the scale height, fully 3-D calculations should be used, such as in \citet{ZL06}. 

In \citet{Tsang2013b}, a companion paper to this work, we utilize the cold disk limit of the results above to explore the effect of stellar illumination on gap opened by a giant planet. We show that disk entropy gradients are sufficiently modified in many cases to allow eccentricity excitation of the planets to occur by reducing the effectiveness of the eccentricity damping corotation resonances. We also suggest that a signature of this process may be evident in the deficit of low-metallicity eccentric planets in the ``Eccentricity Valley'', between $\sim 0.1$ and $\sim 1$au from their host stars \citep{Dawson2013}, as this corresponds to the region that would be shadowed by an inflated dust rim in a low metallicity disk \citep{Dullemond2001, Muzerolle2003}.

\section*{Acknowledgements}
This research was supported by funding from the Lorne Trottier Chair in Astrophysics and Cosmology, and the Canadian Institute for Advanced Research. I would like to thank Andrew Cumming, Kostas Gourgouliatos, Neal Turner, Dong Lai, Peter Goldreich and Francois Foucart for helpful advice and useful discussions during the course of this work.

\appendix
\section{Approximating the Whittaker Function at zero}\label{WhittZero}

Above we have stated that we can approximate ${\rm W}_{\nu,\mu}(i\epsilon) \simeq {\rm W}_{\nu, 1/2}(0) = 1/\Gamma(1-\nu)$, for $\epsilon$ only fairly small. This is not immediately obvious as formally when ${\epsilon \rightarrow 0}$,  ${\rm W}_{\nu,\mu}(i\epsilon) \rightarrow 0$, however it does so only logarithmically slowly, thus only a small imaginary component is sufficient to prevent significant variation from ${\rm W}_{\nu, 1/2}(0)$. This can be seen by first expressing the Whittaker function in terms of the Kummer Function $M(a, b, z)$ \citep{AS65},
\be
{\rm W}_{\nu,\mu}(z) = \frac{\Gamma(-2\mu)}{\Gamma(\tfrac{1}{2} -\mu -\nu)} e^{z/2} z^{\tfrac{1}{2} + \mu} M(\tfrac{1}{2}+\mu - \nu, 1+2\mu, z) + \frac{\Gamma(2\mu)}{\Gamma(\tfrac{1}{2} +\mu -\nu)} e^{z/2} z^{\tfrac{1}{2} - \mu} M(\tfrac{1}{2}-\mu - \nu, 1+2\mu, z).
\ee
As $|z| \rightarrow 0$, $M(a, b, z) = 1$ for $b \notin \mathbb{Z}$. We also have that $0 < \mu < 1/2$, and thus the second term above dominates as $|z|\rightarrow 0$.
Taking $z = i\epsilon$ we have, 
\be
\lim\limits_{\epsilon \rightarrow 0} {\rm W}_{\nu, \mu}(i\epsilon) = \frac{\Gamma(2\mu)}{\Gamma(\tfrac{1}{2} + \mu - \nu)} e^{i\epsilon/2 + i\pi(\tfrac{1}{2} - \mu)/2} \epsilon^{\tfrac{1}{2}- \mu}. \label{Wnearzero}
\ee
As $\epsilon \rightarrow 0$, ${\rm Arg}[{\rm W}_{\nu,\mu}(i\epsilon)] = \pi(\tfrac{1}{2} - \mu)/2+ \epsilon/2$, while
\be
|{\rm W}_{\nu,\mu}(i\epsilon)| = \frac{\Gamma(2\mu)}{\Gamma(\tfrac{1}{2} + \mu - \nu)} \epsilon^{\tfrac{1}{2} - \mu}.
\ee
Thus for small $\epsilon$ we find, expanding in terms of $\tfrac{1}{2}-\mu$,
\be
|{\rm W}_{\nu, \mu}(i\epsilon)| = \frac{\tfrac{1}{2}- \mu}{\Gamma(1 - \nu)} \left(1 + 2\gamma_e + \psi(1-\nu) + \ln \epsilon \right) + {\cal O}[(\tfrac{1}{2} - \mu)^2].
\ee
where $\gamma_e\simeq 0.57722$ is the Euler-Mascheroni constant, and $\psi(z)\equiv \Gamma'(z)/\Gamma(z)$ is the digamma function. 
Defining $\Delta \equiv 1 - |{\rm W}_{\nu,\mu}(i\epsilon)/{\rm W}_{\nu, 1/2}(0)| \ll 1$ to be the fractional error we have from taking ${\rm W}_{\nu,\mu}(i\epsilon) \simeq {\rm W}_{\nu, 1/2}(0) $, we then find that the imaginary component, $\epsilon$, required to have $\Delta$ or less fractional error is given by
\be
\ln \epsilon \go - \frac{\Delta}{\tfrac{1}{2}-\mu} - 2\gamma_e - \psi(1 - \nu) + {\cal O}[\tfrac{1}{2} - \mu],
\ee
therefore the minimum $\epsilon$ required to assume ${\rm W}_{\nu, \mu}(0) \simeq {\rm W}_{\nu, 1/2}(0)$ scales exponentially as $\sim \exp[-\Delta/(\tfrac{1}{2}- \mu)]$, such that only $\epsilon \go 10^{-43}$ is required to have fractional error $\Delta \leq 10\%$, with $\nu = 0.1$ and $\tfrac{1}{2} - \mu = 10^{-3}$.

\renewcommand{\bibsection}{\section{References}} 

\end{document}